\documentclass[12pt]{article}
\usepackage[cp1251]{inputenc}
\usepackage[T2A]{fontenc}
\usepackage[english,russian]{babel}
\usepackage{amsmath,amssymb,amsthm,amscd,latexsym,multicol,array}
\usepackage{graphicx}

\makeatletter

\newcounter{punct}
\newcommand{\punct}[1]{\par
\pagebreak[2]\medskip \refstepcounter{punct} {\bf \thepunct{}.
#1.} }

\def\theequation{
\ifnum\value{punct}<1{\arabic{equation}}
\else{$\!$\thepunct.\arabic{equation}}\fi}
\@addtoreset{equation}{punct}

\renewcommand {\@biblabel}[1]{\hfill #1.}

\newcommand{\code}[1]{\begin{flushleft} УДК #1 \end{flushleft}}

\renewcommand{\maketitle}{\par
\begingroup
   \def\thefootnote{\fnsymbol{footnote}}%
   \def\@makefnmark{\hbox
       to\z@{$\m@th^{\@thefnmark}$\hss}}%
   \if@twocolumn
     \twocolumn[\@maketitle]%
     \else
     \global\@topnum\z@
     \@maketitle \fi\thispagestyle{plain}\@thanks
 \endgroup
 \setcounter{footnote}{0}%
 \let\maketitle\relax
 \let\@maketitle\relax
 \gdef\@thanks{}\gdef\@author{}\gdef\@title{}\let\thanks\relax}

\renewcommand{\@maketitle}{
\begin{center}%
{\@title \par}%
\bigskip
{\large \@author}
\medskip
\end{center}%
}

\renewcommand{\@begintheorem}[2]{\trivlist
  \itemindent=\parindent
  \item[\hskip \labelsep{\bf #1\ #2}]\it}

\begin{document}

\title{СИММЕТРИЯ И ГЕОМЕТРИЧЕСКИ ОБУСЛОВЛЕННЫЕ НЕЛИНЕЙНОСТИ В МЕХАНИКЕ И ТЕОРИИ ПОЛЯ}

\author{Я. Е. Славяновский, В. Ковальчук\\
[5pt]
Институт Фундаментальных Проблем Техники\\ 
Польской Академии Наук,\\
ул. Павиньскего $5^{\rm B}$, 02-106 Варшава, Польша,\\
[5pt]
e-mails: jslawian@ippt.gov.pl, vkoval@ippt.gov.pl}

\code{530.1; 531} 
\setcounter{page}{1}

\maketitle

{\em Рассмотрена взаимосвязь между нелинейностью и симметрией динамических моделей. Особенное внимание уделено существенной (не имеющей характера малых возмущений) нелинейности, когда вообще не существует линейного фона. Такая нелинейность существенно отличается от тех нелинейностей, которые задаются нелинейными поправками, наложенными на некоторый линейный фон. В некотором смысле наши идеи являются продолжением и развитием подхода, положенного Борном и Инфельдом в основу своей электродинамики, а также схем общей теории относительности. Особенно представляют интерес аффинные симметрии степеней свободы и динамические модели. Рассмотрены механические геодезические модели, где упругая динамика тела сосредоточена не в потенциальной энергии, а исключительно в аффинно-инвариантной кинетической энергии, т.е. в аффинно-инвариантных метрических тензорах на конфигурационном пространстве. В некотором смысле это напоминает идею, заключённую в вариационном принципе Мопертюи. Рассмотрена также динамика полей линейных базисов, инвариантная под действием линейной группы внутренних симметрий. Оказалось, что такие модели автоматически имеют структуру обобщённой модели Борна-Инфельда. Этот факт является новым подтверждением идей, впервые предложенных Борном и Инфельдом. Рассмотренные модели могут быть использованы в теории нелинейной упругости и в механике релятивистских сред со структурой. Они также могут привести нас к некоторым альтернативным моделям в теории гравитации. Кроме того существует интересная взаимосвязь между этими моделями и теорией нелинейных интегрируемых цепочек.}

{\bf Ключевые слова:} нелинейность, симметрия, не имеющие характера малых возмущений модели, аффинная инвариантность, нелинейность Борна-Инфельда, аффинно-твёрдые тела, релятивистские среды со структурой, внутренние степени свободы.

\punct{Без нелинейности нет жизни}
Как известно, биологические системы работают в гомеостатических циклических режимах. Они проводят свою жизнь в окрестностях устойчивых притягивающих предельных циклов. Это возможно лишь для объектов, управляемых нелинейными динамическими системами
\begin{equation}\label{eq1}
\frac{dx^{i}}{dt}=f^{i}\left(x^{1},\ldots,x^{n}\right),
\end{equation}
где $i=1,\ldots,n$ и $x^{i}$ --- переменные, описывающие их состояние.

\begin{center}
\includegraphics[scale=0.15]{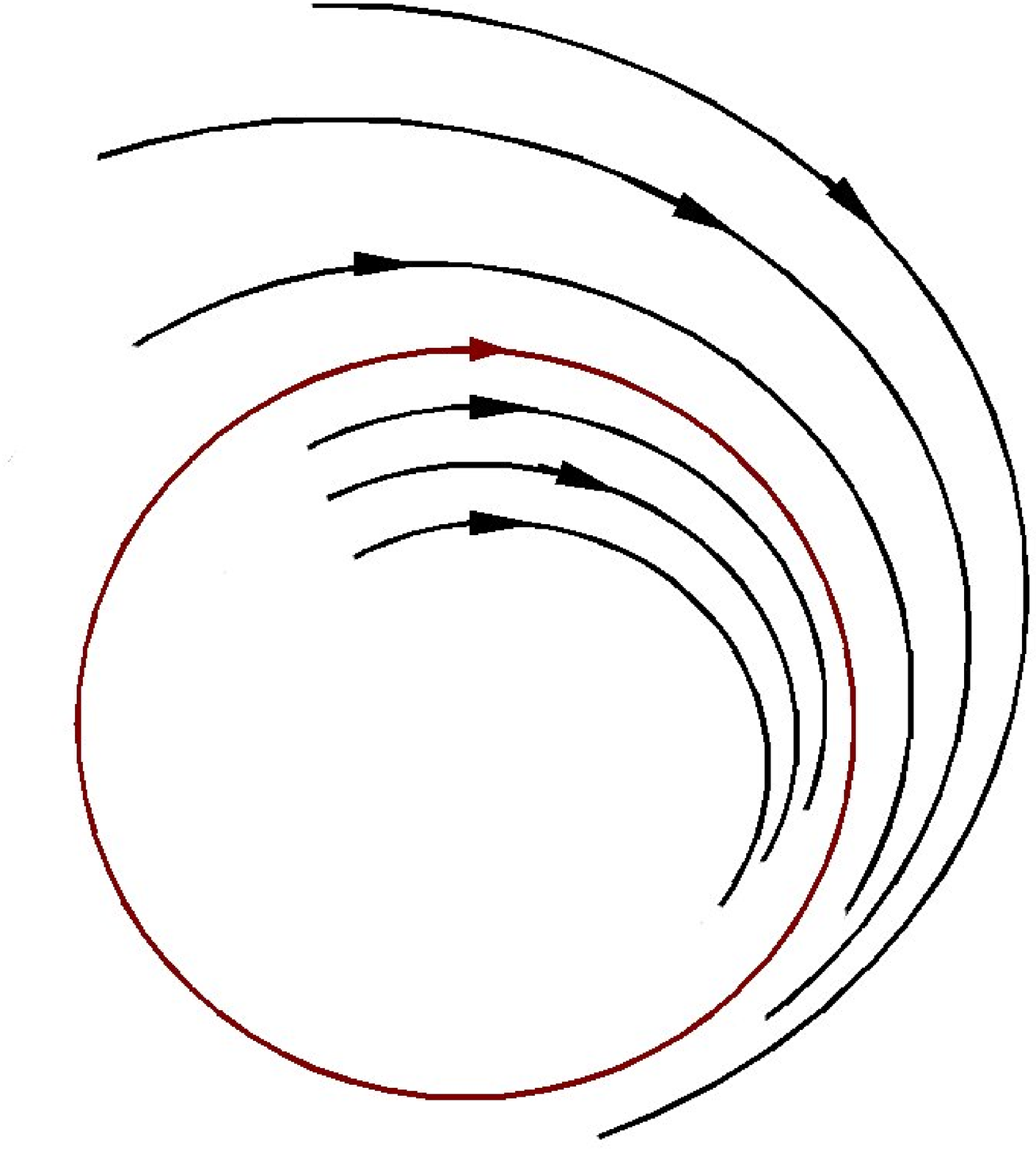}

{\rm Рис. 1}
\end{center}

Другой, быть может тривиальный, но практически очень важный пример --- это тепловое расширение тел. Отталкивающая составляющая межмолекулярного потенциала имеет непараболический вид, обычно с особенностью в $r=0$. Этим объясняется тепловое расширение:

\begin{center}
\includegraphics[scale=0.15]{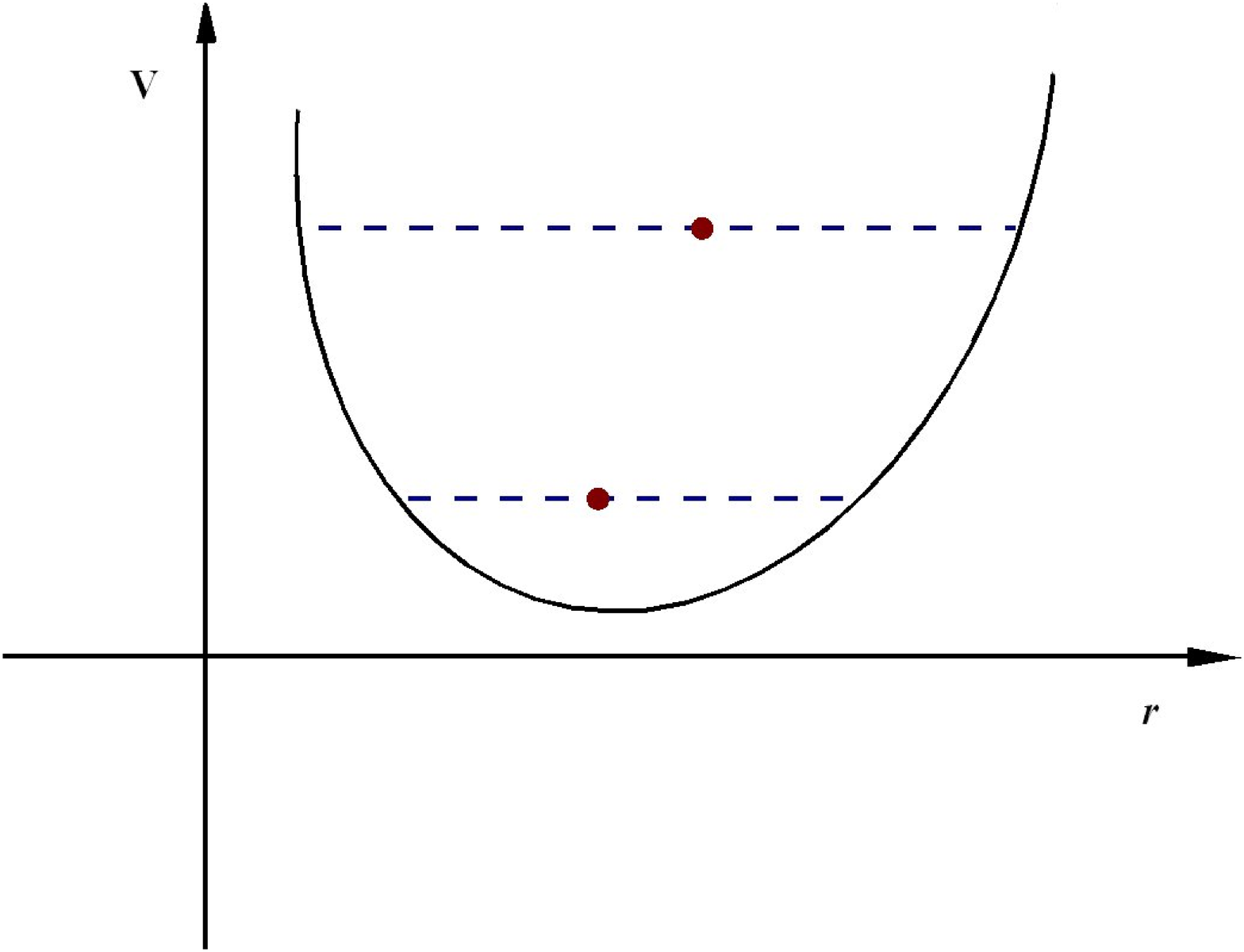}

{\rm Рис. 2}
\end{center}

Линейные гамильтоновы системы являются вполне интегрируемыми, и поэтому в них нет стохастизации энергии, т.е. нет равномерного распределения энергии по степеням свободы.

Все фундаментальные теории поля нелинейны. Неквадратичные слагаемые в их лагранжианах обычно описывают взаимодействие между полями. Системы полей --- элементы некоторого линейного пространства  $H$. В рамках линейных теорий поля уравнения символически могут быть записаны в однородном виде
\begin{equation}\label{eq2}
L\Psi=0
\end{equation}
или, когда заданы определённые внешние источники $f$ (как некоторые элементы пространства $H$), --- в неоднородном виде
\begin{equation}\label{eq3}
L\Psi=f,
\end{equation}
где $L$ --- линейный оператор. 

Как правило, в наиболее употребляемых нелинейных теориях поля уравнения имеют вид
\begin{equation}\label{eq4}
L\Psi+N\left(\varepsilon,\Psi\right)=0,
\end{equation}
где $L$ --- линейная "основа"{}, а сам нелинейный оператор $N$ дополнительно зависит от некоторого параметра  $\varepsilon$. Этот параметр отвечает за нелинейность, так что $N\left(0,\Psi\right)=0$. Обычно предполагается, что линейная задача, т.е. та задача, которая отвечает значению  $\varepsilon=0$, так или иначе "разрешима"{}, и решение нелинейной модели представимо в виде степенных рядов
\begin{equation}\label{eq5}
\Psi=\sum^{\infty}_{n=0}\varepsilon^{n}\Psi_{n}.
\end{equation}
Такая процедура, если и эффективна, то только в тех случаях, когда нелинейный оператор $N$ сам допускает представление в виде степенных рядов,
\begin{equation}\label{eq6}
N\left(\varepsilon,\Psi\right)=
\sum^{\infty}_{n=0}\varepsilon^{n}N_{n}\left(\Psi\right).
\end{equation}
Отправляясь от $\Psi_{0},$ как от чего-то "известного"{}, обычно пытаются "решить"\ иерархию уравнений для $\Psi_{n},$ полагая шаг за шагом равными нулю члены при соответствующих степенях $\varepsilon$. Очевидно, что такая процедура довольно рискованна и ненадёжна --- получающиеся решения имеют символический вид формальных рядов, вопросы сходимости которых к решению исходной задачи требуют дополнительного исследования. В принципе, такая процедура работает для электромагнитных и слабых взаимодействий, но зачастую терпит неудачу при описании сильных взаимодействий. 

В лагранжевых теориях применяют разложения функций Лагранжа
\begin{equation}\label{eq7}
\mathcal{L}\left(\Psi,\partial\Psi\right)=
\sum^{\infty}_{n=0}\varepsilon^{n}\mathcal{L}_{n}\left(\Psi,\partial\Psi\right),
\end{equation}
где для $n>2$ все слагаемые $\mathcal{L}_{n}$ --- многочлены степеней выше чем  $2$ (линейная основа) от аргументов $\left(\Psi,\partial\Psi\right)$. В таких случаях будем говорить, что имеет место нелинейность, имеющая характер возмущений. Во всех остальных случаях будем говорить о существенной нелинейности (СНЛ).

Имеется нечто искусственное и "рукотворное"\ в такого рода нелинейностях, хотя среди них имеются и такие, которые хорошо (а иногда и очень хорошо) подтверждаются экспериментами. Тем не менее, существуют также хорошо определённые модели, в которых нелинейности не имеют характера возмущений. В этих моделях, мотивированных некоторыми идеями симметрии $[1]$, нет какой бы то ни было линейной основы, порождённой квадратичной функцией Лагранжа. Прежде, чем приступить к их обсуждению, лишь для сравнения и лучшего понимания напомним некоторые известные хорошо работающие модели, имеющие характер возмущений.
   
\punct{Системы с взаимодействием определяющего материю комплексного поля Клейна-Гордо\-на и определяющего излучение электромагнитного поля}
Для таких систем лагранжиан имеет следующий вид:
\begin{equation}\label{eq8}
\mathcal{L}=g^{\mu\nu}\overline{D_{\mu}\Psi}D_{\nu}\Psi\sqrt{|g|}-
m^{2}\overline{\Psi}\Psi\sqrt{|g|}-\frac{1}{4}g^{\mu\alpha}g^{\nu\beta}F_{\mu\nu}
F_{\alpha\beta}\sqrt{|g|},
\end{equation}
где $g_{\mu\nu}$ --- ковариантные компоненты метрического тензора, $g$ --- определитель матрицы их коэффициентов, т.е. $g=\det\left[g_{\mu\nu}\right]$, $A_{\mu}$ --- ковектор электромагнитного потенциала, $F_{\mu\nu}$ --- тензор электромагнитного поля,
\begin{equation}\label{eq9}
F_{\mu\nu}=\partial_{\mu}A_{\nu}-\partial_{\nu}A_{\mu},
\end{equation}
и 
\begin{equation}\label{eq9a}
D_{\mu}\Psi=\partial_{\mu}\Psi+ieA_{\mu}\Psi,
\end{equation}
где $e$ --- постоянная взаимодействия (электрический заряд в натуральных единицах). Линейные и квадратичные слагаемые по $e$, отвечающие за нелинейности третьей и четвёртой степеней в функции Лагранжа $\mathcal{L}$ (т.е. квадратичный и кубический вклады в полевые уравнения), соответственно имеют вид
\begin{eqnarray}
&&ieg^{\mu\nu}A_{\mu}\left(\Psi\partial_{\nu}\overline{\Psi}-
\overline{\Psi}\partial_{\nu}\Psi\right)\sqrt{|g|}=g^{\mu\nu}A_{\mu}j_{\nu}=
A_{\mu}j^{\mu},\label{eq10}\\
&&e^{2}g^{\mu\nu}A_{\mu}A_{\nu}\overline{\Psi}\Psi\sqrt{|g|}.\label{eq11}
\end{eqnarray}

\punct{Системы с взаимодействием определяющего излучение поля Максвелла и определяющего фермионную материю поля Дирака}
В этом случае имеем
\begin{eqnarray}
\mathcal{L}&=&\frac{i}{2}e^{\mu}{}_{A}\left(\widetilde{\Psi}\gamma^{A}D_{\mu}
\Psi-\left(D_{\mu}\widetilde{\Psi}\right)\gamma^{A}\Psi\right)\sqrt{|g|}
\nonumber\\
&-&m\widetilde{\Psi}\Psi\sqrt{|g|}-\frac{1}{4}g^{\mu\alpha}g^{\nu\beta}
F_{\mu\nu}F_{\alpha\beta}\sqrt{|g|}.\label{eq12}
\end{eqnarray}
Используемые здесь символы имеют следующий смысл:
\begin{itemize}
\item $e^{\mu}{}_{A}$ --- компоненты неголономного поля реперов, ортонормальных по отношению к метрическому тензору $g$. Их используют для того, чтобы осуществить описание в криволинейных координатах и сделать возможным гладкий переход к искривлённому пространству-времени, так, что
\begin{equation}\label{eq13}
g_{\mu\nu}e^{\mu}{}_{A}e^{\nu}{}_{B}=\eta_{AB},\qquad \left[\eta_{AB}\right]={\rm diag}\left(1,-1,-1,-1\right),
\end{equation}
где $\eta$ --- это плоская метрика Минковского в $\mathbb{R}^{4}$, а не в пространстве-времени.

\item $\gamma^{A}$ --- матрицы Дирака. Они удовлетворяют свойству антикоммутативности
\begin{equation}\label{eq14}
\gamma^{A}\gamma^{B}+\gamma^{B}\gamma^{A}=2\eta^{AB}
\end{equation}
и являются эрмитовыми по отношению к полуторалинейному эрмитову скалярному произведению $G$ нейтральной сигнатуры:
\begin{equation}\label{eq15}
\Gamma^{A}{}_{\overline{r}s}=\overline{\Gamma^{A}{}_{\overline{s}r}}=
G_{\overline{r}z}\gamma^{Az}{}_{s},\quad \left[G_{\overline{r}s}\right]={\rm diag}\left(1,1,-1,-1\right).
\end{equation}

\item сопряжённый биспинор  $\widetilde{\Psi}$ определён как
\begin{equation}\label{eq16}
\widetilde{\Psi}_{r}=\overline{\Psi}^{\overline{s}}G_{\overline{s}r}.
\end{equation}

\item ковариантное дифференцирование биспиноров задаётся соотношением
\begin{equation}\label{eq17}
D_{\mu}\Psi=\partial_{\mu}\Psi+\omega_{\mu}\Psi
\end{equation}
со связностью 
\begin{equation}\label{eq18}
\omega_{\mu}=\frac{1}{8}\Gamma_{KL\mu}=-\frac{1}{8}\Gamma_{LK\mu}=\frac{1}{8}
\eta_{KM}\Gamma^{M}{}_{L\mu}
\end{equation}
а $\Gamma^{K}{}_{L\mu}$ определяется с помощью некоторой $g$-связности Римана-Кар\-тана как
\begin{equation}\label{eq19}
\Gamma^{\alpha}{}_{\beta\mu}=e^{\alpha}{}_{A}\Gamma^{A}{}_{B\mu}e^{B}{}_{\beta}+
e^{\alpha}{}_{A}e^{A}{}_{\beta,\mu},
\end{equation}
так что
\begin{equation}\label{eq20}
\nabla[\Gamma]g=0,
\end{equation}
и $e^{A}{}_{\alpha}$ --- компоненты двойственной котетрады,
\begin{equation}\label{eq21}
e^{A}{}_{\alpha}e^{\alpha}{}_{B}=\delta^{A}{}_{B}.
\end{equation}
\item Лагранжиан кубичен по полевым переменным, полевые уравнения квадратичны по $(A,\Psi)$.
\end{itemize}

В рассмотренных примерах метрический тензор был фиксированной абсолютной величиной. Это сводит группу симметрий к группе изометрий, т.е. преобразований, сохраняющих $g$. Из фиксированности метрики $g$ следует, что нелинейность модели, пусть даже и сильная, не является очень существенной. Причина этого состоит в том, что метрика $g$, необходимая для построения скалярных плотностей из динамических переменных и их производных, сама по себе является внешним элементом, фиксированным раз и навсегда.

\punct{Гравитационная теория как существенно нелиней\-ная модель}
Ситуация полностью меняется, когда мы рассматриваем метрический тензор как динамическую величину, описывающую гравитационное поле. Теперь больше нет объекта, зафиксированного извне, всё должно быть построено из метрики  $g$, и из-за этого теория становится нелинейной по существу, а не за счёт малых возмущений $[2,3]$.

Аналитически метрический тензор представлен симметричным метрическим полем $g_{\mu\nu}\left(x^{\alpha}\right),$ зависящим от пространствен\-но-временных координат. Первая, очень наивная идея построить из метрики $g_{\mu\nu}$ и её производных некоторую, свободную от индексов величину, может выглядеть следующим образом:
\begin{equation}\label{eq22}
g^{\mu\varkappa}g^{\nu\lambda}g^{\alpha\beta}
g_{\mu\nu,\alpha}g_{\varkappa\lambda,\beta}.
\end{equation}
Однако на общем, не обладающем структурой многообразии это выражение будет полностью лишено смысла. Эта величина не является скаляром, также как и величина
\begin{equation}\label{eq23}
g^{\mu\varkappa}g^{\nu\lambda}g^{\alpha\beta}
g_{\mu\nu,\alpha}g_{\varkappa\lambda,\beta}\sqrt{|g|}  
\end{equation}
не является скалярной плотностью. Эйнштейн, будучи мотивирован хорошими физическими идеями, провёл, тем не менее, много лет, борясь с проблемами, подобными данной. Точное решение было найдено Гильбертом, который опирался на своё глубокое понимание математики и римановой геометрии.

Точнее говоря, процедура такова: метрический тензор $g$ порождает связность Леви-Чивита
\begin{equation}\label{eq24}
\left\{\begin{array}{c}
\mu\\
\nu\varkappa
\end{array}\right\}=\frac{1}{2}g^{\mu\alpha}\left(g_{\alpha\nu,\varkappa}+
g_{\alpha\varkappa,\nu}-g_{\nu\varkappa,\alpha}\right).
\end{equation}
Согласно стандартной процедуре, на её основе возникают следующие промежуточные объекты: тензор римановой кривизны  $R^{\alpha}{}_{\beta\mu\nu}$, тензор Риччи $R_{\mu\nu}=R^{\alpha}{}_{\mu\alpha\nu}$ и, наконец, скалярная кривизна $R=g^{\mu\nu}R_{\mu\nu}$. После этого строят скалярную плотность веса один --- величину
\begin{equation}\label{eq25}
R\sqrt{|g|}.
\end{equation}
Эта величина зависит квадратичным образом от первых производных. Слагаемые в  (\ref{eq23}) и (\ref{eq25}), содержащие вторые производные, отличаются на полную дивергенцию. Эта дивергенция не является скалярной плотностью, но она не сказывается на уравнениях поля. Важное положение состоит в том, что в выражениях (\ref{eq23}) и (\ref{eq25}) для свёртки индексов не используются никакие "не-динамические"\ величины. Из-за этого появляются два важных свойства: лагранжиан и результирующие уравнения поля общековариантны, т.е. инвариантны под действием всей группы
\begin{equation}\label{eq26}
{\rm Diff}M\subset {\rm Bij}M
\end{equation}
диффеоморфизмов пространственно-временного многообразия $M$ на себя. Эта крайне высокая симметрия и очень сильная нелинейность, не происходящая из возмущений, соотносятся друг с другом $[2,3]$.

\punct{Модель Эйнштейна-Картана}
Стоит заметить, что здесь допустимо пространственно-вре\-менное многообразие с кручением, или точнее, нет необходимости использовать связность Леви-Чивита. Как раз наоборот, в данных условиях достаточно естественным будет использование модели Эйн\-штейна-Картана. Тогда аффинная связность $\Gamma$ оказывается метрической
\begin{equation}\label{eq27}
\nabla_{\mu}g_{\alpha\beta}=0,
\end{equation}
но необязательно симметричной, т.е. кручение
\begin{equation}\label{eq28}
S^{\lambda}{}_{\mu\nu}=\frac{1}{2}\left(\Gamma^{\lambda}{}_{\mu\nu}-
\Gamma^{\lambda}{}_{\nu\mu}\right)
\end{equation}
не обязано обращаться в нуль. Можно показать, что 
\begin{equation}\label{eq29}
\Gamma^{\lambda}{}_{\mu\nu}=\left\{\begin{array}{c}
\lambda\\
\mu\nu
\end{array}\right\}+
K^{\lambda}{}_{\mu\nu}=\left\{\begin{array}{c}
\lambda\\
\mu\nu
\end{array}\right\}+
S^{\lambda}{}_{\mu\nu}+S_{\mu\nu}{}^{\lambda}+S_{\nu\mu}{}^{\lambda}.
\end{equation}
Вычисляя скалярную кривизну  $R$ этой связности, получим лагранжиан модели Эйнштейна-Картана
\begin{equation}\label{eq30}
L_{\rm EC}=R\left[g,\Gamma\right]\sqrt{|g|},
\end{equation}
где все индексы тензора свёрнуты, подняты или опущены с помощью метрики $g$. Тензорную величину  $K$ обычно называют тензором конторсии. Очевидно модель также общековариантна и существенно нелинейна (СНЛ), так как никакие фиксированные дополнительные объекты не появляются в  $L_{\rm EC}$.

Нужно выделить следующий момент. "Паразитные"\  слагаемые в лагранжиане, содержащем вторые производные от полевых переменных, могут быть удалены инвариантным образом, если в качестве фундаментальных величин вместо метрического тензора использовать поле тетрад (поле линейных базисов) на пространственно-временном многообразии $M$.  Между прочим, это необходимо, если мы хотим включить в общерелятивистские рамки спинорные поля, т.е. фермионную материю. И оказывается, что вследствие этого появляются некоторые новые идеи, касающиеся связи между существенной нелинейностью и инвариантностью. В то же время, такая задача имеет много общего с релятивистской механикой сплошной среды.

\punct{Метод тетрадного поля $[4,5]$}
Пусть $M$ --- $n$-мерное "прос\-транственно-временное"\ многообразие. Специальное значение физической размерности $n=4$ здесь несущественно. Предположим, что многообразие  $M$ параллелизуемо, т.е. допускает гладкое поле линейных базисов (неголономных систем координат).

Главный расслоённый пучок линейных базисов будет обозначен $FM.$ Очевидно, что он представляет собой многообразие размерности $n(n+1)$. Пусть $e=\left(\ldots,e_{A},\ldots\right)$ обозначает поле линейных базисов, т.е. сечение  $FM$ над $M$. Двойственное поле кобазисов будем обозначать следующим образом: $\widetilde{e}=\left(\ldots,e^{A},\ldots\right)$. Если нет опасности, что могут возникнуть какие-то недоразумения, мы не будем различать по написанию величины  $e$ и $\widetilde{e}$ и будем просто использовать аналитические символы $e^{\mu}{}_{A}$, $e^{A}{}_{\mu}$, где
\begin{equation}\label{eq31}
\left\langle e^{A},e_{B}\right\rangle=e^{A}{}_{\mu}e^{\mu}{}_{B}=\delta^{A}{}_{B}.
\end{equation}
Поле $e$ порождает телепараллельную связность $\Gamma_{\rm tel}[e]$, которая определена единственным образом тем условием, что все $e_{A}$ (и автоматически --- $e^{A}$) параллельны,
\begin{equation}\label{eq32a}
\nabla e_{A}=0,\qquad \nabla e^{A}=0.
\end{equation}
Нетрудно показать, что  
\begin{equation}\label{eq32}
\Gamma_{\rm tel}[e]^{\lambda}{}_{\mu\nu}=e^{\lambda}{}_{A}e^{A}{}_{\mu,\nu}.
\end{equation}
Эта связность имеет обращающуюся в нуль кривизну, но её кручение в общем случае ненулевое. Она используется в некоторых задачах теории дислокаций. Параллельный перенос в смысле  $\Gamma_{\rm tel}[e]$ очевидно не зависит от пути. Параллельный перенос некоторого вектора $u$ из точки $a\in M$ в точку $b\in M$ состоит в том, что в точке $b$ берут вектор $v\in T_{b}M,$ который в локально неголономном базисе имеет те же компоненты, что и $u$,
\begin{equation}\label{eq33}
u=x^{A}e_{A}(a),\qquad v=x^{A}e_{A}(b).
\end{equation}
Обсуждаемые ниже тетрадные методы, как уже говорилось, необходимы для общерелятивистской теории спиноров. И кроме того, их можно рассматривать как модель релятивистского континуума с микроструктурой. Метрический тензор в этом описании не является первичной физической величиной, потому что он получается как промежуточная величина из поля линейных базисов $e$,
\begin{equation}\label{eq34}
g=\eta_{AB}e^{A}\otimes e^{B},\qquad \mbox{т.е.}\qquad
g_{\mu\nu}=\eta_{AB}e^{A}{}_{\mu}e^{B}{}_{\nu},
\end{equation}
где $\eta$ --- постоянная метрика Минковского в $\mathbb{R}^{n}$. Тем самым базис  $e$ автоматически $\eta$-ортонормален,
\begin{equation}\label{eq34a}
g\left(e_{A},e_{B}\right)=g_{\mu\nu}e^{\mu}{}_{A}e^{\nu}{}_{B}=\eta_{AB}.
\end{equation}
Группа ${\rm GL}(n,\mathbb{R})$, т.е. структурная группа $FM$, действует естественным образом на $FM$ и $F^{\ast}M$. Это означает, что на многообразиях базисов и ко-базисов для $L\in{\rm GL}(n,\mathbb{R})$ имеем
\begin{eqnarray}
FM\ni e=\left(\ldots,e_{A},\ldots\right)
&\mapsto& eL=\left(\ldots,e_{B}L^{B}{}_{A},\ldots\right),\label{eq35}\\
F^{\ast}M\ni \widetilde{e}=\left(\ldots,e^{A},\ldots\right)
&\mapsto& \widetilde{e}L=\left(\ldots,L^{-1A}{}_{B}e^{B},\ldots\right).\qquad \label{eq35a}
\end{eqnarray} 
Соответствие $e\mapsto g[e]$ является локально-инвариантным под действием (\ref{eq35}), т.е. $L$ может зависеть от точки приложения $e$. Соответствие же $e\mapsto S[e]$ --- глобально инвариантно, т.е. инвариантно под действием постоянной $L,$  независимой от $x^{\mu}$. Лагранжиан Гильберта может быть выражен через  $e$. Введём так называемые инварианты Вейценбока (Weitzenb\"ock)
\begin{equation}\label{eq36}
J_{1}=g_{ia}g^{jb}g^{kc}S^{i}{}_{jk}S^{a}{}_{bc},\ \ J_{2}=g^{ij}S^{k}{}_{li}S^{l}{}_{kj},\ \ J_{3}=g^{ij}S^{a}{}_{ai}S^{b}{}_{bj}.
\end{equation}
Они квадратичны по производным  $e$ (потому что $S$ линейно по $de$). После некоторых вычислений можно показать, что лагранжиан Гильберта может быть выражен как
\begin{equation}\label{eq37}
R[g]\sqrt{|g|}=
\left(J_{1}+2J_{2}-4J_{3}\right)\sqrt{|g|}+4\nabla_{i}\left(
S^{a}{}_{ab}g^{bi}\sqrt{|g|}\right).
\end{equation}
Но связность Леви-Чивита симметрична, а последнее выражение --- ковариантная дивергенция некоторой векторной плотности веса один. Поэтому эта величина --- обычная дивергенция, и её можно отбросить без изменения уравнений движения. Модифицированный лагранжиан Гильберта принимает вид
\begin{equation}\label{eq38}
L^{\prime}\left(e,\partial e\right)=\left(J_{1}+2J_{2}-4J_{3}\right)\sqrt{|g|}.
\end{equation}
Выражения  (\ref{eq36}) глобально инвариантны относительно (\ref{eq35}), а  соотношение (\ref{eq38}) локально инвариантно по модулю несущественных дивергентных поправок. И очевидно, что выражения (\ref{eq36}), (\ref{eq38}) также общековариантны, т.е. ${\rm Diff}M$-инвариантны, в точности, как гильбертов лагранжиан. И теперь возникает вопрос: почему не допустить в (\ref{eq38}), т.е. в  $L^{\prime}$, некоторые более общие соотношения между коэффициентами, нежели частное отношение $1:2:(-4)$ $[4$--$9]$? Такая замена не нарушает общей ковариантности. Верно то, что согласно (\ref{eq35}) она нарушает локальную инвариантность под действием ${\rm GL}(n,\mathbb{R})$. Но глобальная инвариантность под действием внутренней группы Лоренца  ${\rm O}(n,\eta)$ остаётся справедливой. Имеются некоторые признаки того, что такие модифицированные модели жизнеспособны. Общая ковариантность гораздо более важна, так что, быть может, принятие предположения о ней, означающее также отказ от локальной инвариантности под действием внутренней группы ${\rm O}(n,\eta)\subset{\rm GL}(n,\mathbb{R}),$ делает упор на принятии вместо этого предположения об инвариантности под действием глобальной группы  ${\rm GL}(n,\mathbb{R}),$ в то время как локальная инвариантность, очевидно, невозможна. Хотя целая ${\rm GL}(n,\mathbb{R})$, действующая как (\ref{eq35}), описывает основные свойства геометрии степеней свободы в $FM$ $[4$--$9]$.

\punct{Модели типа Борна-Инфельда}
Следуя и дальше этим путём, мы найдём довольно неожиданную и новую мотивировку для очень интересных классов обобщённых моделей типа Борна-Инфельда. На самом деле, простейший класс моделей, общековариантных и глобально инвариантных под действием  ${\rm GL}(n,\mathbb{R})$ имеет следующий вид
\begin{equation}\label{eq39}
L=\sqrt{\left|\det\left[L_{\mu\nu}\right]\right|},
\end{equation}
где $L_{\mu\nu}$, т.е. компоненты так называемого тензора Лагранжа, заданы как
\begin{equation}\label{eq40}
L_{\mu\nu}=AS^{\alpha}{}_{\beta\mu}S^{\beta}{}_{\alpha\nu}+
BS^{\alpha}{}_{\alpha\mu}S^{\beta}{}_{\beta\nu}+
CS^{\beta}{}_{\beta\alpha}S^{\alpha}{}_{\mu\nu},
\end{equation}
где $A$, $B$, $C$ --- некоторые постоянные $[9]$.

Это наиболее общий тензор, построенный как квадратичная функция от производных $\partial e$ вектора $e$ посредством квадратичной зависимости от величины $S$, и, в то же время, построенный из поля  $e$ согласно общековариантному правилу в $M$ и аморфным, ${\rm GL}(n,\mathbb{R})$-инвариантным способом во внутреннем пространстве $\mathbb{R}^{n}$. 

Его симметричная часть, т.е. комбинация первых двух слагаемых
\begin{equation}\label{eq41}
T_{\mu\nu}=AS^{\alpha}{}_{\beta\mu}S^{\beta}{}_{\alpha\nu}+
BS^{\alpha}{}_{\alpha\mu}S^{\beta}{}_{\beta\nu},
\end{equation}
как можно ожидать, играет роль метрического тензора. Для сравнения напомним, что правило Дирака-Эйнштейна (\ref{eq34})
\begin{equation}\label{eq41a}
\eta_{AB}e^{A}{}_{\mu}e^{B}{}_{\nu}
\end{equation}
также было общековариантным в $M$ и локально ${\rm O}(n,\eta)$-инва\-риантным в  $\mathbb{R}^{n}$. Эта последняя симметрия, будучи бесконечномерной, в этом смысле богаче чем та, что задаётся соотношением (\ref{eq41}). Однако она беднее в смысле неподчинения симметрии ${\rm GL}(n,\mathbb{R})$. Эта последняя симметрия в определённом смысле более фундаментальна для  $FM$ и не предусматривает никакого фиксированного объекта во внутреннем пространстве $\mathbb{R}^{n}$. Особый интерес представляет само первое слагаемое в (\ref{eq41}), в то время как остальные играют просто роль вторичных поправок. Все дело в том, что выражение
\begin{equation}\label{eq42}
g[e]_{\mu\nu}=S^{\alpha}{}_{\beta\mu}S^{\beta}{}_{\alpha\nu}
\end{equation}
построено из $S$ в соответствии с алгебраической конструкцией Киллинга. Это совпадение не случайно. Пусть $\Omega^{A}{}_{BC}$ обозначает объект неголономии на $e$,
\begin{equation}\label{eq43}
\left[e_{A},e_{B}\right]=\Omega^{C}{}_{AB}e_{C},\qquad de^{A}=\frac{1}{2}\Omega^{A}{}_{BC}e^{C}\wedge e^{B}.
\end{equation}
С точностью до постоянного множителя это $e$-неголономное представление самого $e$, т.е.
\begin{equation}\label{eq44}
S^{\lambda}{}_{\mu\nu}=\frac{1}{2}\Omega^{C}{}_{AB}e^{\lambda}{}_{C}
e^{A}{}_{\mu}e^{B}{}_{\nu},
\end{equation}
или, записанное иными словами с помощью  $M$-индексов,
\begin{equation}\label{eq45}
S=\frac{1}{2}\Omega^{C}{}_{AB}e_{C}\otimes e^{A}\otimes e^{B}.
\end{equation}
Можно показать, что всякий лагранжиан, построенный из $e$ общековариантным и ${\rm GL}(n,\mathbb{R})$-инвариантным образом, должен зависеть от $S$, и поэтому также от производных $\partial e$ однородным образом степени однородности  $n$. Единственный способ рассмотреть общий случай состоит в том, что надо ввести некоторые скалярные выражения построенные из $g[e]$ (\ref{eq42}) и $S[e]$. Можно показать, что такие скаляры, всегда однородные степени однородности ноль в $S$ (или в $\partial e$), не вносят ничего качественно нового, но усложняют формулу для $L\left(e,\partial e\right)=L(S)$. Если поле базисов оказывается натянутым на алгебру Ли, т.е. если $\Omega$ постоянна, то $(M,e)$ становится пространством группы Ли, и $g[e]$ пропорционально её тензору Киллинга:
\begin{equation}\label{eq46}
g[e]_{\mu\nu}=4\gamma[e]_{AB}e^{A}{}_{\mu}e^{B}{}_{\nu},\qquad \gamma[e]_{AB}=\Omega^{K}{}_{LA}\Omega^{L}{}_{KB},
\end{equation}
где $\Omega^{A}{}_{BC}$ --- не что иное, как структурные постоянные.

Важный и приводящий в сильное волнение вывод состоит в том, что если алгебра Ли полупроста, то $e$ --- всегда решение полевых уравнений, получающихся из лагранжиана $L$ как уравнения Эйлера-Лагранжа. Эти решения называются "однородными"\ или "вакуумными"\ решениями.

\punct{Взаимосвязь между существенными нелинейностями, не имеющими характера малых возмущений, и группами симметрии высокого порядка}
Представленная выше модель была построена для некоторых, очень специальных типов геометрических объектов и основывалась на некоторой, очень специальной мотивировке. Однако эта мотивировка оказалась весьма глубокой. Она как раз и выражает нашу идею о существовании глубокой и неотъемлемой связи между существенной нелинейностью, не имеющей характера малых возмущений, и группами симметрии высокого порядка (быть может, скрытыми).

Лагранжианы вида (\ref{eq39}), (\ref{eq40}) очень интересны как альтернативный полюс простоты в смысле дополнительности к лагранжианам, квадратичным по скоростям (и приводящим к квазилинейным уравнениям поля). Они были впервые обнаружены в нелинейной электродинамике (как модели Борна-Инфельда). Но мотивировка тогда была совершенно другой, ориентированной на то, чтобы избежать трудностей, присущих линейной электродинамике Максвелла. Характерное свойство линейных и квазилинейных теорий поля состоит в том, что лагранжианы квадратичны по производным. Кажется, что эта возможность --- наиболее простая. Но можно найти и другой, альтернативный путь, имеющий глубокое основание в геометрии, симметрии и нелинейности. Действительно, геометрическая структура всякого лагранжиана  $\mathcal{L}\left(\Psi,\partial\Psi\right)$ --- это структура скалярной плотности веса один (или, более точно, плотность Вейля, $W$-плотность) на пространственно-временном многообразии, или, более общо, на "многообразии независимых переменных". В полевых теориях специальной теории относительности, или, в более общем случае, в теориях, сформулированных на основе некоторого фиксированного или динамического метрического тензора, скажем, в общей теории относительности, имеется стандартная процедура построения таких величин, т.е.
\begin{equation}\label{eq41b}
\mathcal{L}\left[g,\Psi\right]=L\left[g,\Psi\right]\sqrt{|g|},
\end{equation}
где $L\left[g,\Psi\right]$ --- скалярная функция, построенная из полей, и $\sqrt{|g|}$, т.е. определитель метрического тензора, --- это стандартная скалярная плотность веса один. Затем, когда основываясь на такой факторизации, будет довольно естественным ожидать, что простейшие и наиболее эффективные модели линейны и квазилинейны, когда $L$ зависит от производных полиномиально, и полином имеет вторую (или по крайней мере низкого порядка) степень. Однако, если мы однажды примем философию скалярных плотностей как что-то фундаментальное, тогда, также как и в случае (\ref{eq39}) мы окажемся склонными верить в противоположный полюс простоты. Имеется в виду, что не сам лагранжиан, но некоторый дважды ковариантный тензор Лагранжа $L_{\mu\nu}\left(\Psi,\partial\Psi\right)$ должен быть рассмотрен как нечто фундаментальное, потому что извлечение квадратного корня из (абсолютного значения) ковариантного тензора второго порядка --- это канонический способ построения лагранжевых плотностей и вариационных принципов. Поэтому не лагранжиан, а именно, тензор Лагранжа $L_{\mu\nu}\left(\Psi,\partial\Psi\right)$ должен иметь как можно более простую форму, например как в (\ref{eq39}), (\ref{eq40}). Таким образом, ожидается, что $L_{\mu\nu}$ будет некоторым полиномом второго (или по крайней мере низкого) порядка по $\partial\Psi$. 

Эта идея очень стара и восходит ещё к некоторым попыткам Борна и Инфельда избежать противоречий в классической электродинамике. Эти попытки почти полностью забыты, но недавно интерес к этой модели возродился вновь $[1,10]$.

\punct{Традиционная модель Борна-Инфельда в электродинамике} 
Пусть $A_{\mu}$ --- 4-потенциал электромагнитного поля. Тогда интенсивность поля задана с помощью кососимметричного тензора
\begin{equation}\label{eq41c}
F_{\mu\nu}=\partial_{\mu}A_{\nu}-\partial_{\nu}A_{\mu},
\end{equation}
т.е. внешней производной (в этом случае, грубо говоря, операцией взятия ротора)
от $A_{\mu}$. Основные инварианты можно представить в следующем виде:
\begin{eqnarray}
S&=&-\frac{1}{4}F_{\mu\nu}F^{\mu\nu}=-\frac{1}{4}g^{\mu\varkappa}g^{\nu\lambda}
F_{\mu\nu}F_{\varkappa\lambda}=\frac{1}{2}\left(\overline{E}^{2}-
\overline{B}^{2}\right),\quad\\
P&=&-\frac{1}{4}F_{\mu\nu}\check{F}^{\mu\nu}=-\frac{1}{8}
\varepsilon^{\mu\nu\lambda\varkappa}F_{\mu\nu}F_{\varkappa\lambda}=
\overline{E}\cdot\overline{B},
\end{eqnarray}
где  $S$ --- скаляр, $P$ --- псевдоскаляр, а $\overline{E}$ и
$\overline{B}$ --- соответственно электрическое поле и магнитная индукция. Компоненты вектора $\overline{E}$, т.е. $E_{i}$, равны  компонентам $F_{0i}$ (или $F_{i0}=-F_{0i}$, в зависимости от конвенции) тензора $F$, а компоненты псевдовектора $\overline{B}$, т.е. $B_{i}$, совпадают с компонентами $F_{jk}=-F_{kj}$, $j\neq k$, $k\neq i$, $i\neq j$. Следует обратить внимание читателя на то, что мы следуем здесь релятивистской конвенции, где греческие индексы являются пространственно-временными, а латинские --- только пространственными. Таким образом $\mu$, $\nu$ пробегают целый диапазон $0,1,2,3$, в то время как $i$, $j$, $k$ --- только $1,2,3$. Кроме того в вышеприведённых формулах предполагается также суммирование по повторяющимся индексам.

Пусть $L[F]=\ell(S,P)$ --- лагранжиан, а $g_{\mu\nu}$ --- метрика Минковского. Далее мы будем использовать псевдо-ортогональные координаты, так что
\begin{equation}\label{eq41d}
\left[g_{\mu\nu}\right]={\rm diag}\left(1,-1,-1,-1\right).
\end{equation}

Традиционный лагранжиан Максвелла в линейной электродинамике имеет вид 
\begin{equation}\label{eq41e}
L=S.
\end{equation}

Хорошо известный лагранжиан в нелинейной электродинамике Борна-Инфельда записывается как
\begin{eqnarray}\label{eq46B}
L&=&b^{2}-b^{2}\sqrt{1-\frac{2}{b^{2}}S-\frac{1}{b^{4}}P^{2}}\nonumber\\
&=&
b^{2}\sqrt{\left|\det\left[g_{\mu\nu}\right]\right|}-
\sqrt{\left|\det\left[bg_{\mu\nu}+F_{\mu\nu}\right]\right|}.
\end{eqnarray}
При этом вышеприведённая модель была сформулирована уже в окончательной версии теории; в начале же предложенная Борном оригинальная модель основывалась на лагранжиане 
\begin{equation}\label{eq46C}
L=b^{2}\left(\sqrt{1+\frac{1}{b^{2}}\left(\overline{B}^{2}-
\overline{E}^{2}\right)}-1\right).
\end{equation}
Обе модели (\ref{eq46B}) и (\ref{eq46C}) дают тот же результат, когда мы ищем стационарные, сферически симметричные решения полевых уравнений. Однако существуют и некоторые различия в других предсказаниях этих двух теорий.

Основная мотивировка для поисков нелинейных моделей в электродинамике состоит в следующем:
\begin{itemize}
\item получить конечную собственную электромагнитную энергию электрона (а следовательно и его конечную электромагнитную массу), а возможно и эффект насыщения (сатурации), т.е. максимальную силу электростатического поля,
\item иметь возможность получить уравнения движения для заряженной частицы из уравнений поля, например как для массивных частиц в общей теории относительности $[2,3,11]$.
\end{itemize}
Идеальной схемой в этом случае является схема появления существенной нелинейности в общей теории относительности, т.е. в релятивистской теории гравитации Эйнштейна.

Сферически симметричные стационарные решения полевых уравнений, которые мы получаем для моделей (\ref{eq46B}) и (\ref{eq46C}), явно удовлетворяют вышеприведённым требованиям. Если мы потребуем, чтобы электростатический потенциал $\varphi=A_{0}$ ($A_{i}=0$) был функцией только от радиальной переменной $r$, тогда мы в конечном счёте получим следующую картину, которая иллюстрирует качественные различия между электродинамикой Максвелла и Борна-Инфельда:

\begin{center}
\includegraphics[scale=0.25]{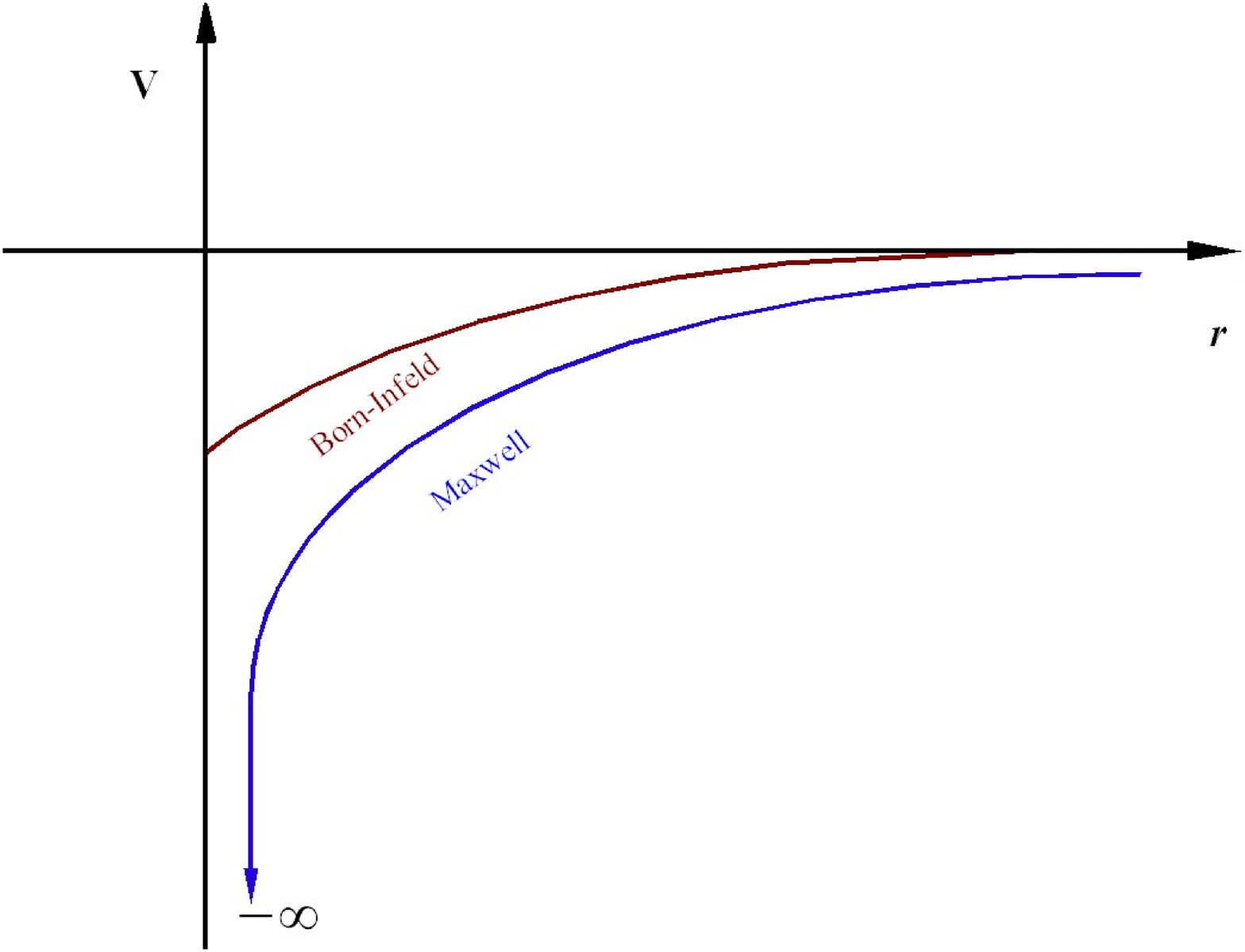}

{\rm Рис. 3}
\end{center}

Соответствующее аналитическое описание имеет вид:
\begin{equation}\label{eq41f}
\overline{E}(\overline{r})=\frac{e}{\sqrt{r^{4}_{0}+x^{4}}}
\frac{\overline{r}}{r},\quad \varphi(r)=\int^{\infty}_{r}\frac{edx}{\sqrt{r^{4}_{0}+x^{4}}},\quad r_{0}=\sqrt{\frac{e}{b}},
\end{equation}
где $e$ --- это постоянная интегрирования, которая с физической точки зрения тождественна электрическому заряду электрона, а $\overline{r}$ обозначает радиус-вектор.

Легко видеть, что скалярный потенциал $\varphi$ конечен, электрическое поле $\overline{E}$ ограничено (хотя и неопределено в точке $r=0$), плотность энергии $\omega$, т.е. компонента $T_{00}$ тензора энергии-импульса, бесконечна в точке $r=0$, но тем не менее общая энергия (т.е. масса электрона) $\mathcal{E}=\int\omega d_{3}\overline{r}$ конечна.

\punct{Исключительность модели Борна-Инфельда}
В свое время имели место многочисленные попытки построения нелинейной электродинамики, но среди всех них модель Борна-Инфельда является в определённом смысле каноничной и исключительной в силу следующих своих свойств:   
\begin{itemize}
\item она калибровочно инвариантна,

\item функционал энергии является положительно определённым,

\item точечные источники имеют конечную электромагнитную массу,

\item в релятивистском пространстве-времени поток энергии не является пространственно-подобным вектором,

\item в ней нет двойного лучепреломления,

\item в ней существуют плоские волны на фоне постоянного электромагнитного поля, а также уединённые волны.
\end{itemize}
Вследствие этого теория Борна-Инфельда казалась наиболее интересной среди всех возможных нелинейных моделей электродинамики, но, тем не менее, в конечном счёте она оказалась явно разочаровывающей по следующим причинам:
\begin{itemize}
\item не было убедительных результатов в ожидаемой импликации: уравнения поля $\Rightarrow$ уравнения движения. Успех общей теории относительности в этом отношении опирался не только на наличии нелинейности, хотя и был с нею сильно связан. Главным образом это произошло благодаря тождествам Бьянки, которые следуют из общей ковариантности. Существенная нелинейность была здесь конечно важна, но не напрямую, а только как следствие общей ковариантности.

\item не было похоже на то, что спектры сверхтяжёлых атомов подтверждают идеи Борна-Инфельда,

\item имелись серьёзные трудности при переходе к квантовой версии теории обусловленные неполиномиальной структурой лагранжиана,

\item одной из её главных мотивировок была бесконечность электромагнитной массы в линейной теории Максвелла. Но вскоре появилась квантовая электродинамика, и благодаря процедуре ренормализации учёные перестали больше бояться бесконечностей. Иными словами, предполагалось, что масса электрона не имеет только электромагнитного происхождения, вследствие этого эти две бесконечности взаимно сокращаются и масса получается конечной.

\item буквально говоря, историческая модель Борна-Инфельда не очень хорошо подходит при описании "внешней"\ заряженной материи (отличной от "внутренней"{}, которая описывается особенностями поля магнитной индукции $\overline{D}$). Например, для квантовой когерентной материи мы имеем
\begin{eqnarray}\label{eq41g}
L&=&b^{2}\sqrt{|g|}-\sqrt{bg+F}\nonumber\\
&+&g^{\mu\nu}\overline{D_{\mu}\Psi}D_{\nu}\Psi\sqrt{|g|} 
-m^{2}\overline{\Psi}\Psi\sqrt{|g|},
\end{eqnarray}
где $D_{\mu}=\partial_{\mu}+ieA_{\mu}$. Возникающие здесь полевые уравнения имели бы нерациональную структуру в полевых переменных и их производных.
\end{itemize}

Вследствие вышеприведённых причин интерес к модели Борна-Инфель\-да после некоторого времени совсем исчез с физического "рынка"{}. Время от времени модель была использована как квазиклассическое описание некоторых нелинейных явлений, таких как, например, рассеяние фотонов на фотонах. В результате её предсказания совпадали с теми, которые были получены с помощью квантовой электродинамики, но процедура была во многих аспектах значительно проще. Нелинейное взаимодействие между линейным полем Максвелла и материей было в этом случае закодировано в нелинейности Борна-Инфельда в виде чисто электромагнитной модели. Грубо говоря, нелинейность полевой динамики замещает материю, эффективно её описывая.

\punct{Аффинно-твёрдые тела}
Теория аффинно-твёрдых тел представляет собой в определённом смысле механический пример, лежащий в основе нашей идеи ${\rm GL}(n,\mathbb{R})$-инвариант\-ной гравитации, основанной на линейных базисах $[12$--$20]$.

\begin{center}
\includegraphics[scale=0.3]{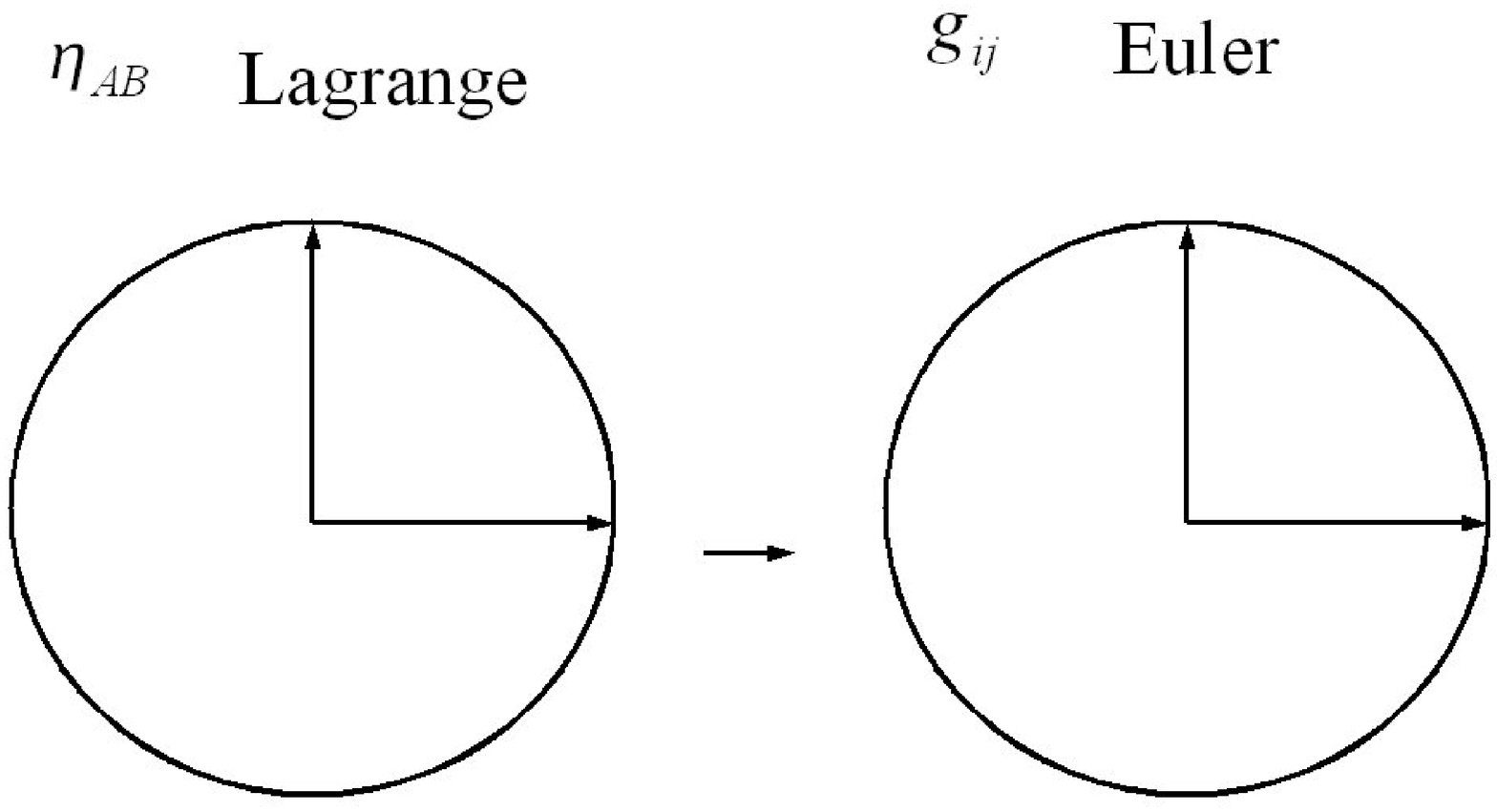}

{\rm Рис. 4}
\end{center}

Во время движения все аффинные соотношения между составляющими тела сохраняются. В общем это не так с метрическими соотношениями. Поэтому лагранжевы и эйлеровы координаты $a^{K}$, $\xi^{i}$ связаны между собой как 
\begin{equation}\label{eq41h}
\xi^{i}(t,a)=x^{i}+\varphi^{i}{}_{K}(t)a^{K},
\end{equation}
где $x^{i}$ --- положение центра масс, а $\varphi^{i}{}_{K}$ --- координаты внутреннего движения. В $n$-мерном пространстве в совокупности имеются  $n(n+1)$ обобщённых координат $x^{i}$ и $\varphi^{i}{}_{K}$.

Мы имеем обобщённые скорости $v^{i}$, $V^{i}{}_{K}$ и некоторые другие определяемые ими величины, такие как аффинные скорости $\Omega$, $\widehat{\Omega}$ в пространственном и связанном с телом представлениях 
\begin{eqnarray}
v^{i}&=&\frac{dx^{i}}{dt},\quad \widehat{v}^{A}=\varphi^{-1A}{}_{i}v^{i},
\quad V^{i}{}_{K}=\frac{d\varphi^{i}{}_{K}}{dt},\label{eq43a}\\
\Omega^{i}{}_{j}&=&V^{i}{}_{K}\varphi^{-1K}{}_{j}=
\frac{d\varphi^{i}{}_{K}}{dt}\varphi^{-1K}{}_{j},\label{eq43b}\\
\widehat{\Omega}^{A}{}_{B}&=&\varphi^{-1A}{}_{i}V^{i}{}_{B}=
\varphi^{-1A}{}_{i}\frac{d\varphi^{i}{}_{B}}{dt}=
\varphi^{-1A}{}_{i}\Omega^{i}{}_{j}\varphi^{j}{}_{B}.\label{eq43c}
\end{eqnarray}
Они представляют собой аффинные аналоги угловых скоростей из механики твёрдого тела.

\punct{"Обычная"\ кинетическая энергия $T$, основанная на принципе Даламбера}
В этом случае имеем функцию Лагранжа
\begin{equation}\label{eq44a}
L=T-\mathcal{V}(x,\varphi),
\end{equation}
где $\mathcal{V}(x,\varphi)$ --- это потенциальная энергия, кинетическая энергия записывается как
\begin{equation}\label{eq44b}
T=\frac{m}{2}g_{ij}\frac{dx^{i}}{dt}\frac{dx^{j}}{dt}+
\frac{1}{2}g_{ij}\frac{d\varphi^{i}{}_{A}}{dt}\frac{d\varphi^{j}{}_{B}}{dt}
J^{AB}=T_{\rm tr}+T_{\rm int},
\end{equation}
а величины
\begin{equation}\label{eq44c}
m=\int d\mu(a),\qquad J^{AB}=\int a^{A}a^{B}d\mu(a) 
\end{equation}
соответственно описывают трансляционную (масса) и внутреннюю инертность. Здесь $T$ --- это кинетическая энергия описывающая линейный фон.

\punct{Динамически аффинно-инвариантные модели. Связь между сильной нелинейностью и симметрией высокого порядка}
Вышеприведённая модель, которая хотя и основана на аффинных степенях свободы, не является аффинно-инвариантной на уровне динамики, так что ее симметрия слишком бедна. Однако мы можем ввести модели, аффинно-инвариантные в физическом пространстве:
\begin{eqnarray}
{}^{\rm left}T_{\rm int}&=&\frac{1}{2}\mathcal{L}^{B}{}_{A}{}^{D}{}_{C}
\widehat{\Omega}^{A}{}_{B}\widehat{\Omega}^{C}{}_{D},\label{eq47a}\\
{}^{\rm left}T_{\rm tr}&=&\frac{m}{2}\eta_{AB}
\widehat{v}^{A}\widehat{v}^{B}=\frac{m}{2}C_{ij}
v^{i}v^{j},\label{eq47b}
\end{eqnarray}
где $C_{ij}=\eta_{AB}\varphi^{-1A}{}_{i}\varphi^{-1B}{}_{j}$ --- тензор деформаций Коши, и материальном:
\begin{eqnarray}
{}^{\rm right}T_{\rm int}&=&\frac{1}{2}\mathcal{R}^{j}{}_{i}{}^{l}{}_{k}
\Omega^{i}{}_{j}\Omega^{k}{}_{l},\label{eq47c}\\
{}^{\rm right}T_{\rm tr}&=&\frac{m}{2}g_{ij}
v^{i}v^{j}=\frac{m}{2}G_{AB}\widehat{v}^{A}\widehat{v}^{B},\label{eq47d}
\end{eqnarray}
где $G_{AB}=g_{ij}\varphi^{i}{}_{A}\varphi^{j}{}_{B}$ --- тензор деформаций Грина.

Здесь $\mathcal{L}$, $\mathcal{R}$ --- некоторые постоянные. Но $\Omega$ и $\widehat{\Omega}$ --- неголономные скорости, поэтому метрики, лежащие в основе этих $T$, являются неевклидовыми, и даже без потенциала динамика инвариантна относительно "большой"\ аффинной группы, не только относительно сохраняющих метрику групп (изометрий) как в разд. 12.

Модели, одновременно инвариантных относительно левых и правых (пространственных и материальных) действий аффинных групп, не существует. Но сама внутренняя кинетическая энергия может быть двусторонне аффинно-инвариант\-на. Однако такие кинетические энергии знаконеопределены:
\begin{equation}\label{eq53}
{}^{\rm aff}T^{\rm aff}_{\rm int}=\frac{A}{2}{\rm Tr}\left(\Omega^{2}\right)+
\frac{B}{2}\left({\rm Tr}\Omega\right)^{2}=\frac{A}{2}{\rm Tr}\left(\widehat{\Omega}^{2}\right)+
\frac{B}{2}\left({\rm Tr}\widehat{\Omega}\right)^{2}.
\end{equation}

Кроме того, существуют интересные модели, аффинно-ин\-вариантные в физическом пространстве и только изометрически инвариантные в материальном пространстве, и наоборот. Это соответственно
\begin{eqnarray}
{}^{\rm aff}T^{\rm metr}_{\rm int}&=&\frac{I}{2}\eta_{AC}\eta^{BD}\widehat{\Omega}^{A}{}_{B}
\widehat{\Omega}^{C}{}_{D}
+{}^{\rm aff}T^{\rm aff}_{\rm int},\label{eq54}\\
{}^{\rm metr}T^{\rm aff}_{\rm int}&=&\frac{I}{2}g_{ik}g^{jl}\Omega^{i}{}_{j}
\Omega^{k}{}_{l}+{}^{\rm aff}T^{\rm aff}_{\rm int},\label{eq55}
\end{eqnarray}
где $I$, $A$ и $B$ --- постоянные, описывающие инерционные свойства тела во время внутреннего (относительного) движения. Другими словами это инерция по отношению к оборотам и однородным деформациям.

Если величины $I$, $A$ и $B$ удовлетворяют некоторым ограничениям, тогда выражения (\ref{eq54}), (\ref{eq55}) могут быть положительно определены. Многообразие таких троек $(I,A,B)$ --- это открытое подмножество в $\mathbb{R}^{3}$.

Следуя вышеописанной схеме, которая задана выражениями (\ref{eq47a})--(\ref{eq47d}), можно построить некоторые интересные риманновские структуры на многообразии скалярных произведений, а точнее, на многообразии (обычно симметричных) билинейных форм заданных на действительном линейном пространстве и на многообразии (обычно эрмитовых) полуторалинейных форм заданных на комплексном линейном пространстве. Вышеупомянутые риманновские структуры являются также инвариантными под действием "большых"\ групп, например целой линейной группы как в случае моделей (\ref{eq47a})--(\ref{eq55}). Кроме того, получающиеся уравнения для их геодезических кривых также являются очень сильно, существенно нелинейными.

Существует надежда, что такие модели нелинейной динамики на многообразиях полуторалинейных скалярных произведений могут быть полезны при описании оснований теории квантов. Возможно также, что они могут быть выгодны в объяснении проблемы декогеренции и квантовых парадоксов измерений.

В каждом случае предложенная нелинейность является геометрической, натуральной и не имеет характера малых возмущений. С уверенностью можно также сказать, что она не введена "вручную"{}, как в некоторых искусственных моделях нелинейности в квантовой механике. 

\punct{Сильно нелинейные реалистические геодезические модели, их строгая разрешимость и цепочки}
Среди всех этих аффинных моделей с кинетической энергией, аффинно-инвариантной как минимум в физическом или материальном пространстве (\ref{eq53})--(\ref{eq55}) имеются такие, которые задают упругую динамику лишь в самой кинетической энергии, безо всяких потенциалов. Это имеет место как минимум в том случае, когда  $\varphi$ сохраняет объём. Такие модели могут описывать упругие колебания, даже если они чисто геодезические, т.е. не содержат дополнительного потенциала. Это напоминает принцип Мопертюи. И такие модели, по крайней мере частично, могут быть явно решены на основе некоторых экспоненциальных матричных функций. Всё это имеет место благодаря замечательной связи между существенной нелинейностью и симметриями высших порядков. Это напоминает наши аффинно-инвариантные модели, формально определённые для полей линейных базисов в гравитации. 

Связь с интегрируемыми цепочками можно увидеть, если воспользоваться так называемым биполярным разложением
\begin{equation}\label{eq45a}
\varphi=LDR^{-1},
\end{equation}
где матрицы  $L$ и $R$ ортогональны, а матрица $D$ диагональна. 

Тогда $L$ и $R$ описывают два некоторых фиктивных гироскопа (ортонормальные главные оси тензоров деформаций Коши и Грина). Тогда вводят сопутствующие угловые скорости этих гироскопов,
\begin{equation}\label{eq45b}
\widehat{\chi}=L^{-1}\frac{dL}{dt}=-\widehat{\chi},\qquad \widehat{\vartheta}=R^{-1}\frac{dR}{dt}=-\widehat{\vartheta},
\end{equation}
и инварианты деформаций, т.е. диагональные элементы матрицы $D$:
\begin{equation}\label{eq45c}
D={\rm diag}\left(Q^{1},\ldots,Q^{n}\right).
\end{equation}
Удобно воспользоваться представлением
\begin{equation}\label{eq45d}
D_{ii}=Q^{i}=\exp{q^{i}},
\end{equation}
т.е.
\begin{equation}\label{eq45bd}
q^{i}=\ln{Q^{i}}.
\end{equation}
Далее целесообразно ввести концепцию угловых моментов (спинов) для гироскопов $L$ и $R$, которые определяются как величины, канонически сопряжённые к $\widehat{\chi}$, $\widehat{\vartheta}$. Эти канонические спины обозначим  $\widehat{\varrho}=-\widehat{\varrho}$ и $\widehat{\tau}=-\widehat{\tau}$ соответственно. Введём также следующие вспомогательные величины:
\begin{equation}\label{eq45e}
M=-\widehat{\varrho}-\widehat{\tau},\qquad N=\widehat{\varrho}-\widehat{\tau},
\end{equation}
которые, конечно, кососимметричны. Канонические моменты, сопряжённые  $q^{i},$ будут обозначены через $p_{i},$ а моменты, сопряжённые переменным  $Q^{i}$ --- через $P_{i}$.

Тогда можно показать, что в рамках модели (\ref{eq53}), рассмотренной для простоты в случае, когда $B=0$ (это на самом деле только вторичная поправка), имеем
\begin{equation}\label{eq56}
{}^{\rm aff}\mathcal{T}^{\rm aff}_{\rm int}=\frac{1}{2\alpha}C(2),
\end{equation}
где $\alpha$ --- множитель инерционного происхождения, а $C(2)$ --- вторая функция Казимира:
\begin{eqnarray}\label{eq56a}
C(2)&=&\sum_{a}p^{2}_{a}+
\frac{1}{16}\sum_{a,b}\frac{(M^{a}{}_{b})^{2}}{{\rm sh}^{2}(q^{a}-q^{b})/2}\nonumber\\
&-&
\frac{1}{16}\sum_{a,b}\frac{(N^{a}{}_{b})^{2}}{{\rm ch}^{2}(q^{a}-q^{b})/2}.
\end{eqnarray}
Второе слагаемое описывает взаимное отталкивание инвариантов деформации, а третье --- их притяжение. Вследствие этого имеются области ограниченного и рассеивающего движения (и существует чёткая граница между ними) даже в чисто геодезических моделях (предполагая только их аффинную инвариантность). Другими словами получаем геодезическое моделирование колебаний --- конечно же за исключением дилатаций, которые стабилизируются с помощью некоторого потенциала $\mathcal{V}(q)$. Общие изотропные потенциалы имеют вид: 
\begin{equation}\label{eq58a}
\mathcal{V}=\mathcal{V}\left(q^{1},\ldots,q^{n}\right).
\end{equation}

Дополнительно, принимая модель из разд. 12, упрощённую за счёт подстановки 
\begin{equation}\label{eq58}
J^{AB}=I\delta^{AB},
\end{equation}
выражающей изотропию тензора инерции, получаем
\begin{eqnarray}\label{eq57}
\mathcal{T}_{\rm int}&=&\frac{1}{2I}\sum_{a}P^{2}_{a}+
\frac{1}{8I}\sum_{a,b}\frac{(M^{a}{}_{b})^{2}}{(Q^{a}-Q^{b})^{2}}\nonumber\\
&+&
\frac{1}{8I}\sum_{a,b}\frac{(N^{a}{}_{b})^{2}}{(Q^{a}+Q^{b})^{2}}.
\end{eqnarray}
Ясно, что формулы (\ref{eq56}), (\ref{eq56a}), (\ref{eq57}) демонстрируют некоторое глубокое родство с интегрируемыми гиперболическими цепочками Сазерлэнда и Калоджеро $[21]$ и могут быть использованы для анализа их решений. Отрицательный вклад в (\ref{eq56a}) отвечает за тот факт, что соответствующие геодезические модели могут описывать устойчивые, в каком-то смысле "ограниченные"{}, упругие колебания.
\vspace{1cm}

\noindent {\em Работа выполнена в рамках программы сотрудничества между Польской и Российской Академиями Наук.}
\bigskip

\centerline{\large ЛИТЕРАТУРА }
\begin{enumerate}
\item
{\em S\l awianowski J.J.} Internal geometry, general covariance and generalized Born-Infeld models. Part I. Scalar fields // Arch.\ Mech. 1994. Vol. 46. No.~3. P. 375--397.

\item 
{\em Misner C.M., Thorne K.S. and Wheeler J.A.} Gravitation. San Francisco: Freeman. 1973.

\item 
{\em Bergmann P.G.} Introduction to the Theory of Relativity. New York: Prentice-Hall. 1942. 

\item
{\em M\"oller C.K.} Title //  Mat.\ Fys.\ Skr.\ Dan.\ Vid.\ Selsk. 1978. Vol. 39. No.~13. P. 1.

\item
{\em Pellegrini C. and Pleba\'{n}ski J.} Title // Mat.\ Fys.\ Skr.\ Dan.\ Vid.\ Selsk. 1963. T. 2. No.~4. P. 1.

\item
{\em S\l awianowski J.J.}$\:$ Generally-covariant field theories and space-time as a micromorphic continuum. Warsaw: Prace IPPT --- IFTR Reports. 1988. Vol. 51. 92 p.

\item
{\em S\l awianowski J.J.} Search for fundamental models with affine symmetry: some results, some hypotheses and some essay // In: Proceedings of the Sixth International Conference on Geometry, Integrability and Quantiza\-tion (June 3-10, 2004, Varna, Bulgaria). Editors: Iva\"{\i}lo M. Mladenov and Allen C. Hirshfeld. Sofia: SOFTEX. 2005. P. 126--172.

\item
{\em S\l awianowski J.J.} Geometrically implied nonlinearities in me\-chanics and field theory // In: Proceedings of the Eighth In\-ternational Conference on Geometry, Integrability and Quan\-tization (June 9-14, 2006, Varna, Bulgaria). Editors: Iva\"{\i}lo M. Mladenov and Manuel de Le\'{o}n. Sofia: SOFTEX. 2007. P. 48--118.

\item
{\em S\l awianowski J.J.} ${\rm GL}(n,\mathbb{R})$ as a candidate for fundamental symmetry in field theory // Il\ Nuovo\ Cimento\ B. 1991. Vol. 106. P. 645--668.

\item 
{\em Chru\'{s}ci\'{n}ski D. and Kijowski J.} Generation of multipole moments by external field in Born-Infeld non-linear electro\-dynamics // J.\ Phys.\ A: Math.\ Gen. 1998. Vol. 31. P. 269--276.

\item 
{\em Infeld L. and Pleba\'{n}ski J.} Motion and Relativity. Oxford: Pergamon. 1960.

\item
{\em Burov A.A. and Chevallier D.P.} Dynamics of affinely de\-formable bodies from the standpoint of theoretical mechanics and differential geometry // Rep.\ on Math.\ Phys. 2008. Vol. 62. No.~3. P. 325--363.

\item
{\em Capriz G.} Continua with Microstructure // Springer Tracts in Natural Philosophy. Vol. 35. New York-Berlin-Heidelberg-Paris-Tokyo: Springer-Verlag. 1989.

\item
{\em Capriz G. and Mariano P.M.} Symmetries and Hamiltonian Formalism for Complex Materials // Journal of Elasticity. 2003. Vol. 72. P. 57--70.

\item
{\em Chevallier D.P.} On the Foundations of Ordinary and Gene\-ralized Rigid Body Dynamics and the Principle of Objectivity // Arch.\ Mech. 2004. Vol. 56. No.~4. P. 313--353.

\item
{\em S\l awianowski J.J., Kovalchuk V., S\l awianowska A., Go\l ubow\-ska B., Mar\-tens A., Ro\.zko E.E. and Zawistowski Z.J.} Inva\-riant geodetic systems on Lie groups and affine models of internal and collective degrees of freedom. Warsaw: Prace IPPT --- IFTR Reports. 2004. Vol. 7. 164 p.

\item
{\em S\l awianowski J.J., Kovalchuk V., Go\l ubowska B., Martens A. and Ro\.zko E.E.} Dynamical systems with internal degrees of freedom in non-Euclidean spaces. Warsaw: Prace IPPT --- IFTR Reports. 2006. Vol. 8. 129 p.

\item
{\em S\l awianowski J.J., Kovalchuk V., S\l awianowska A., Go\l ubow\-ska B., Mar\-tens A., Ro\.z\-ko E.E. and Zawistowski Z.J.} Affine symmetry in mechanics of collective and internal modes. Part I. Classical models // Rep.\ on Math.\ Phys. 2004. Vol. 54. No.~3. P. 373--427.

\item
{\em S\l awianowski J.J., Kovalchuk V., S\l awianowska A., Go\l ubow\-ska B., Mar\-tens A., Ro\.z\-ko E.E. and Zawistowski Z.J.} Affine symmetry in mechanics of collective and internal modes. Part II. Quantum models // Rep.\ on Math.\ Phys. 2005. Vol. 55. No.~1. P. 1--45.

\item
{\em S\l awianowski J.J. and S\l awianowska A.K.} Virial coefficients, collective modes and problems with the Galerkin procedure //  Arch.\ Mech. 1993. Vol. 45. No.~3. P. 305--331.

\item 
{\em Calogero F. and Marchioro C.J.} Exact solution of a one-dimensional three-body scattering problem with two-body and/or three-body inverse-square potentials // Math.\ Phys.\ 1974. Vol. 15. P. 1425.

\end{enumerate}


\code{530.1; 531} 

Симметрия и геометрически-обусловленные нелинейности в механике и теории поля. Славяновский Я. Е., Ковальчук В.

Symmetries and geometrically implied nonlinearities in mecha\-nics and field theory. Warsaw (Poland): Institute of Fundamental Technological Research of Polish Academy of Sciences. 2009. P. ???--???.

Рассмотрена взаимосвязь между нелинейностью и симметрией динамических моделей. Особенное внимание уделено существенной (не имеющей характера малых возмущений) нелинейности, когда вообще не существует линейного фона. Такая нелинейность существенно отличается от тех нелинейностей, которые задаются нелинейными поправками, наложенными на некоторый линейный фон. В некотором смысле наши идеи являются продолжением и развитием подхода, положенного Борном и Инфельдом в основу своей электродинамики, а также схем общей теории относительности. Особенно представляют интерес аффинные симметрии степеней свободы и динамические модели. Рассмотрены механические геодезические модели, где упругая динамика тела сосредоточена не в потенциальной энергии, а исключительно в аффинно-инвариантной кинетической энергии, т.е. в аффинно-инвариантных метрических тензорах на конфигурационном пространстве. В некотором смысле это напоминает идею, заключённую в вариационном принципе Мопертюи. Рассмотрена также динамика полей линейных базисов, инвариантная под действием линейной группы внутренних симметрий. Оказалось, что такие модели автоматически имеют структуру обобщённой модели Борна-Инфельда. Этот факт является новым подтверждением идей, впервые предложенных Борном и Инфельдом. Рассмотренные модели могут быть использованы в теории нелинейной упругости и в механике релятивистских сред со структурой. Они также могут привести нас к некоторым альтернативным моделям в теории гравитации. Кроме того существует интересная взаимосвязь между этими моделями и теорией нелинейных интегрируемых цепочек.

Discussed is relationship between nonlinearity and symmetry of dynamical models. The special stress is laid on essential, non-perturbative nonlinearity, when none linear background does exist. This is nonlinearity essentially different from ones given by nonli\-near corrections imposed onto some linear background. In a sense our ideas follow and develop those underlying Born-Infeld electro\-dynamics and general relativity. We are particularly interested in affine symmetry of degrees of freedom and dynamical models. Discussed are mechanical geodetic models where the elastic dyna\-mics of the body is not encoded in potential energy but rather in affinely-invariant kinetic energy, i.e., in affinely-invariant metric tensors on the configuration space. In a sense this resembles the idea of Maupertuis variational principle. We discuss also the dynamics of the field of linear frames, invariant under the action of linear group of internal symmetries. It turns out that such models have automatically the generalized Born-Infeld structure. This is some new justification of Born-Infeld ideas. The suggested models may be applied in nonlinear elasticity and in mechanics of relativistic continua with microstructure. They provide also some alternative models of gravitation theory. There exists also some interesting relationship with the theory of nonlinear integrable lattices.

{\bf Keywords:} nonlinearity, symmetry, non-perturbative models, affine invariance, Born-Infeld nonlinearity, affinely-rigid bodies, relativistic structured continua, internal degrees of freedom.

\end{document}